\documentclass[aps,prl,twocolumn]{revtex4-1}
\usepackage{graphicx}
\usepackage[english]{babel}
\usepackage{amssymb}
\usepackage{amsbsy}
\usepackage{amsmath}
\usepackage{bbold}
\usepackage{bm}
\usepackage{refcount}

\newcommand{\der}{\mathrm{d}}

\begin{document}
\title{Chemical Continuous Time Random Walks}
\author{Tom\'{a}s Aquino}
\email[e-mail: ]{tomas.aquino@idaea.csic.es}
\affiliation{Spanish National Research Council (IDAEA--CSIC), 08034 Barcelona, Spain}
\author{Marco Dentz}
\affiliation{Spanish National Research Council (IDAEA--CSIC), 08034 Barcelona, Spain}

\begin{abstract}
Kinetic Monte Carlo methods such as the Gillespie algorithm model
chemical reactions as random walks in particle number space. The
inter-reaction times are exponentially distributed under the
assumption that the system is well mixed.
We introduce an arbitrary inter-reaction time distribution, which
may account for the impact of incomplete mixing on chemical
reactions, and in general stochastic reaction delay, which may represent the impact of extrinsic noise. This process
defines an inhomogeneous continuous time random walk in particle number space, from which we derive a 
generalized chemical master equation. This leads naturally to a generalization of the Gillespie algorithm.
Based on this formalism, we determine the modified chemical rate laws for different inter-reaction time
distributions. This framework traces Michaelis--Menten-type kinetics back to finite-mean delay times, and predicts
 time-nonlocal macroscopic reaction kinetics as a consequence of broadly distributed delays. Non-Markovian kinetics exhibit weak ergodicity breaking and show key features of reactions under local non-equilibrium. 
\end{abstract}

\maketitle


Chemical reactions are the result of the interaction between different system components. 
Classically, it is assumed that within a given support volume reactants are well mixed. In other words, all reactants are 
equally available to react at a constant rate. In this case, inter-reaction times due to intrinsic stochastic variability can be shown to be exponentially distributed~\cite{Doob1942,gillespie1977exact}. These observations form the basis of Kinetic Monte Carlo (KMC) methods,
such as the Gillespie algorithm~\cite{gillespie1977exact}, which comprise an important class of models and techniques for the stochastic simulation of reactive systems and population dynamics in general~\cite{chatterjee2007overview,hammersley2013monte}.
The probability distribution of chemical species numbers follows Markovian dynamics in time, which are described by the classical chemical master equation. 
The corresponding macroscopic dynamics are the familiar local rate laws for species concentrations~\cite{gillespie1992rigorous,van1992stochastic}. 
Since chemical reactions are essentially contact processes leading to nonlinear dynamics, this type of framework finds broad application in population dynamics, modeling scenarios as varied as biological cellular processes, disease spread in epidemiology, dynamics on and of networks, animal species interactions in ecology, quantum molecular dynamics, and chemical reactions in geological media~\cite{barkai2000biological,lewis2003autoinhibition,williams2000simple,wieland2012structure,grossman2005efficient,nightingale1998quantum,Steefel2005}. 

Complex dynamics in heterogeneous environments may manifest themselves in terms of effective, distributed delay times affecting the reaction processes. Transport processes are often at the core of non-Poissonian reaction dynamics, since they are the limiting factor on reactant mixing~\cite{Avraham2005,Tartakovsky2008,Battiato:2009,benichou2010geometry}. Medium heterogeneity may affect the efficiency of tracer particles in exploring their surroundings~\cite{condamin2007first,guigas2008sampling,meroz2011distribution,DRG2016PRE}, thus leading to broad distributions of inter-reaction times or reaction rate constants~\cite{Gowri2007}. Furthermore, the nonlinear character of reactions may lead to the amplification of local concentration fluctuations, enhancing the effects of transport limitations and significantly slowing down reactions~\cite{ovchinnikov1978role}. Heterogeneity and fluctuation processes not inherent to the chemical reaction itself are referred to as 
extrinsic noise. Modeling the impact of extrinsic noise on chemical reactions in the KMC sense requires a framework capable of representing more complex inter-reaction times, which describe for example transport-induced delays or unresolved reaction sequences~\cite{hethcote1995sis,bratsun2005delay,Barrio2006,Cai2007,brett2013stochastic}. 

The classical chemical master equation rests on two pillars: Exponential waiting times between reactions, and statistical equivalence of all particles of a given species. The present work removes the first assumption and thereby implicitly relaxes the second, 
providing a unified theoretical framework to quantify the impact of arbitrary inter-reaction times. The continuous time random walk (CTRW) provides a systematic starting point to account for general waiting time distributions between reaction events~\cite{silver1977monte,scher1975anomalous,montroll1973random,scher1973stochastic,berkowitz2006modeling}.
Building from CTRW theory, we derive a generalized chemical master equation capable of accounting for non-exponential inter-reaction times and the resulting non-Markovian character of reaction dynamics in time. In the KMC spirit, the dynamics are represented in terms of a random walk in particle number space rather than in physical space. To the best of our knowledge, this letter provides the first instance of a generalized chemical master equation for a KMC framework that does not assume Markovian (i.e., exponential) waiting times. This allows us to rigorously describe the effects of intrinsic and extrinsic variability of the waiting times, and make corresponding predictions about the large-scale behavior. Our approach derives Michaelis--Menten-type kinetics as a result of random delay times with finite mean, and predicts time-nonlocal macroscopic reaction kinetics as a consequence of broadly distributed delays. The latter show weak ergodicity breaking, a fingerprint of anomalous transport~\cite{klages2008anomalous,bel2005weak,burov2011single,lomholt2007subdiffusion,weigel2011ergodic,jeon2011vivo}, and exhibit key features of 
local non-equilibrium such as power law mass decay.


\paragraph{Framework --}
In order to cast the dynamics of $m_s$ different species that participate in $m_r$ different reactions
into a CTRW framework, we first define the state space. The chemical species are denoted by $\mathcal S_j$, where $j = 1,\dots,m_s$; the corresponding particle numbers are denoted by $n_j$. The state vector of particle numbers is $\bm{n} = (n_1,\ldots,n_{m_s})^\top$, where the superscript $\top$ denotes the transpose. During a reaction $i$ the loss (gain) in particle number $n_j$ is denoted by $r_{ij}\in\mathbb{N}$ ($p_{ij} \in \mathbb N$). These coefficients are typically, but need not be, given by the law of mass action. Thus, the impact of reaction $i$ on the state space can be expressed as
%
\begin{align}
\label{eq::reactions}
	\sum_j r_{ij} \mathcal{S}_j \rightarrow \sum_j p_{ij} \mathcal{S}_j. 
\end{align}
%
The stoichiometric coefficients $s_{ij}=p_{ij} - r_{ij}$ denote the net change in each species $j$ due to each reaction $i$. A single event of reaction $i$ is characterized by the reaction waiting time $\tau^r_{(i)}$ whose probability density function (PDF) $\psi_i^r$ depends in general on the system state $\bm n$; we will elaborate on its specific form below. The reaction event that actually occurs is the one whose waiting time is minimum. Thus, the waiting time between reaction events is $\tau^r = \min\{\tau_{(i)}^r|i =1,\dots,m_r\}$. The joint distribution $\phi_i^r(t;\bm n) \der t$ of reaction $i$ happening and the reaction waiting time being in $[t,t+\der t]$ is then given by (see Appendix~\ref{a::waiting})
%
\begin{align}
\label{eq::tau_reaction_pdf}
	\phi^r_i(t;\bm{n}) = \psi^r_i(t;\bm{n}) \prod_{\ell \neq i}\int_t^\infty \der t_\ell \, \psi_\ell^r(t_\ell;\bm{n}),
\end{align}
%
which states that $\phi_i^r(t;\bm n)$ is given by the probability that the reaction times of the $\ell \neq i$ reactions are larger than the one for reaction $i$, multiplied by the PDF of the waiting time of reaction $i$, $\psi_i^r$.

For the modeling of system fluctuations in terms of waiting times, we distinguish between intrinsic and extrinsic noise. Extrinsic noise results from external fluctuations, that is, variability in the physical or chemical environment. Under transport-limited conditions, reaction delays arise from mass transfer limitations due to reactants' spatial sampling efficiency and fluctuation-induced segregation~\cite{ovchinnikov1978role,benichou2010geometry}. In the KMC spirit, these delays affect all particles in the same way independently of the system state. This is in contrast to intrinsic noise, which by definition represents the inherent stochasticity of the reaction process proper~\cite{van1992stochastic,bratsun2005delay,swain2002intrinsic}. Thus, we introduce a global delay time $\tau^g$ such that for a given state $\bm n$ the inter-reaction time is $\tau = \tau^r(\bm n) + \tau^g(\tau^r)$. The global delay does not depend directly on the state, but may depend on the current reaction waiting time $\tau^r$. As mentioned above, $\tau^g$ is a manifestation of extrinsic noise, and the reaction waiting times $\tau^r$ of intrinsic noise. The joint distribution for reaction $i$ to happen after an inter-reaction time in $[t,t+\der t]$ is denoted by $\phi_i(t,\bm n) \der t$. We consider two global delay scenarios. {\em Scenario 1} assumes that $\tau^g$ is independent of the reaction-specific waiting times and identically distributed, with density $\psi^g$. In this case, we have $\phi_i(t;\bm{n}) = (\phi^r_i * \psi^g)(t;\bm{n})$, where $*$ denotes convolution. {\em Scenario 2} considers $\tau^g$ to be given by a compound Poisson process as $\tau^g(\tau^r) = \sum_{k=1}^{\eta(\tau^r)} \vartheta^{g}_{k}$, where $\eta(u)$ is Poisson-distributed with mean $\gamma u$; the density of the identical independently distributed $\vartheta_k^{g}$ is denoted by $\psi^g_0$. The joint distribution $\phi_i(t;\bm n)$ can be expressed in Laplace space as $\tilde\phi_i(\lambda;\bm{n}) = \tilde\phi^r_i(\lambda + \gamma [ 1 - \tilde\psi^g_0(\lambda) ];\bm{n})$~\cite{Margolin:et:al:2003,comolli2016non}. Laplace transformed quantities are denoted by a tilde, and the Laplace variable is denoted by $\lambda$. Both scenarios represent global delays of the full reaction system. In scenario $1$ the delay is synchronized with the reaction events themselves. The delay time can be seen as a global ``preparation'' time for the next reaction event. This means the delay time is external but the delay event is triggered by the reaction event. In scenario $2$ both delay time and occurrence of delay events (characterized by the rate $\gamma$) are prescribed externally. Such fixed-rate delay events can be related to fluctuation-induced spatial segregation~\cite[][]{Lapeyre2017}.

The CTRW dynamics for the stochastic process describing the random particle number vector $\bm{N}_k$ and time $T_k$ after $k$ reaction steps can now be defined by the recursion relations
%
\begin{align}
\label{eq::ctrw}
	\bm{N}_{k+1} = \bm{N}_k + \bm{s}_{r_k} \;, &&
	\qquad T_{k+1} = T_k + \tau_{k}, 
\end{align}
%
where $\bm s_{r_k} = (s_{r_k1},\dots,s_{r_km_s})^\top$ and the random number $r_k \in (1,\dots,m_r)$ indicates the reaction that is occurring. The joint distribution of $(r_k,\tau_k)$ is given by $\phi_i(t;\bm n)$. 
The initial conditions are deterministic, $\bm{N}_0 = \bm n_0$ and $T_0=0$. The recursion relations~\eqref{eq::ctrw} define an inhomogeneous multi-dimensional CTRW because the joint distribution of $(r_k,\tau_{k})$ depends on the current system state $\bm N_k$. 
We use the CTRW formalism~\cite[][]{KMS73} to derive the following generalized chemical master equation for the probability $P(\bm n,t)$ of finding the system in state $\bm n$ at time $t$ (see Appendix~\ref{a::meq}),
%
\begin{align}
	\partial_t P(\bm{n},t) &= \sum_i \int_0^t \der t' \left(\prod_j \mathbb{E}_j^{-s_{ij}}-1\right) 
\nonumber\\
&\qquad\qquad \times
P(\bm{n},t') M_i(t-t';\bm{n}) \;,
\label{eq::meq}
\end{align}
%
where the step operator $\mathbb{E}^z_j$ acts on a function $f(\bm n)$ by incrementing the particle number $n_j$ of species $\mathcal S_j$ by the integer $z \in \mathbb{Z}$, i.e.,  $\mathbb{E}^z_j f(\bm n) = f(n_1,\dots,n_j + z,\dots,n_{m_s})$ \cite{van1992stochastic}. The memory functions $M_i$ are defined by their Laplace transforms as (see Appendix~\ref{a::meq})
%
\begin{align}
\label{eq::memory}
	\tilde M_i(\lambda;\bm{n}) = \frac{\lambda \tilde \phi_i(\lambda;\bm{n})}{1-\sum_\ell \tilde\phi_\ell(\lambda;\bm{n})},
\end{align}
%
whose form is typical of the CTRW key formalism~\cite[][]{KMS73}. Note that~\eqref{eq::meq} describes the full evolution of the non-linear dynamic system~\eqref{eq::ctrw}, in which the random increments depend on the system state. The generalized chemical master equation is inhomogeneous in that the memory function depends explicitly on the state vector $\bm n$. It generalizes the chemical master equation \cite[][]{gillespie1977exact, gillespie1992rigorous}. A generalized Gillespie algorithm corresponding to~\eqref{eq::meq} is described in Appendix~\ref{a::gillespie}.



\paragraph{Chemical rate laws --}
In order to characterize the impact of stochastic delay on macroscopic reaction dynamics, we focus on the corresponding rate laws. The dimensionless concentrations are defined by $\bm{C} = \bm{N} / n_0$, with $n_0 = \sum_j n_{j,0}$. The macroscopic concentration is given by the ensemble average $\langle \bm{C} \rangle$. We derive the following macroscopic equations (see Appendix~\ref{a::rate_laws}),
%
\begin{equation}
\label{eq::rate_law}
	\partial_t \langle\bm{C} \rangle = \sum_i \bm{s}_i \int_0^\infty \der t' \, \langle M^C_i[t-t';\bm{C}(t')] \rangle \;,
\end{equation}
%
where we define $M^C_i[t;\bm{C}(t)] = M_i[t;n_0 \bm{C}(t)] / n_0$. Note that these key equations are in general not closed. Nontrivial scenarios for which closures of~\eqref{eq::rate_law} are available, and situations for which they are not, are discussed in the following. 



\paragraph{Reaction waiting times --}
The waiting time associated with reaction $i$ events is given by the minimum intrinsic reaction time, which is distributed according to a given PDF $p_i$. Thus, the state-dependent density of waiting times for reaction $i$ is (see Appendix~\ref{a::waiting})
%
\begin{equation}
\label{eq::mass_action}
	\psi_i^r(t;\bm{n}) = h_i(\bm{n})p_i(t)\left[\int_t^\infty \der t' \, p_i(t')\right]^{h_i(\bm{n})-1} \;,
\end{equation}
%
where $h_i(\bm{n})=\prod_j n_j!/[r_{ij}!(n_j-r_{ij})!]$ accounts for all possible combinations of necessary reactants. 
Thus, we obtain from~\eqref{eq::tau_reaction_pdf} for the joint density that reaction $i$ happens with the reaction time $t$
%
\begin{equation}
\label{eq::phi_mass_action}
	\phi^r_i(t;\bm{n}) = \frac{h_i(\bm{n})p_i(t)}{\int_t^\infty \der t' \, p_i(t')} \prod_{\ell=1}^{m_r} \left[\int_t^\infty \der t_\ell \, p_\ell(t_\ell)\right]^{h_\ell(\bm{n})} \;.
\end{equation}
In the following, we briefly discuss the intrinsic reaction waiting time statistics before we analyze in detail the impact of reaction delay due 
to extrinsic noise. 

\paragraph{Intrinsic reaction waiting times --} The intrinsic reaction waiting times are a consequence of the intrinsic system noise. In the proposed KMC framework, the intrinsic waiting times are reset after a reaction event. Considering the reaction process as a superposition of renewal processes~\cite[][]{Cox1954,boguna2014simulating}, this implies that the time to the next reaction after a certain time has elapsed, that is, the forward recurrence time, has the same distribution as the reaction time itself. This is a property of the exponential distribution only. Thus, in the following, we consider the intrinsic reaction waiting times to be exponentially distributed. For $p_i(t) = \kappa_i e^{-\kappa_i t}$, with $\kappa_i$ the (microscopic) reaction rate, the joint distribution~\eqref{eq::phi_mass_action} becomes $\phi_i^r = h_i \kappa_i \exp(-\sum_\ell \kappa_\ell h_\ell t)$. In the absence of delay, that is, for $\phi_i \equiv \phi_i^r$, the memory function is obtained by Laplace inversion of~\eqref{eq::memory} as $M_i = h_i \kappa_i \delta(t)$, where $\delta(\cdot)$ is the Dirac delta. The generalized chemical master equation~\eqref{eq::meq} then becomes the well-known chemical master equation~\cite{gillespie1992rigorous}, which describes Markovian dynamics. The kinetic rate laws are obtained from~\eqref{eq::rate_law} by approximating $h_i(\bm{n}) \approx \prod_j n_j^{r_{ij}} / r_{ij}!$ for large $n_j$ as 
%
\begin{equation}
\label{rate:eq}
	\partial_t \langle\bm{C} \rangle = \sum_i \bm{s}_i \kappa^C_i \prod_j \langle C_j \rangle^{r_{ij}},
\end{equation}
%
where $\langle C_j^{r_{ij}} \rangle \approx \langle C_j \rangle^{r_{ij}}$ for large particle numbers because $P$ becomes strongly peaked about the ensemble average~\cite{van1992stochastic}. The (macroscopic) rate constants are given by $\kappa_i^C = n_0^{\alpha_i-1} \kappa_i / \prod_j r_{ij}!$, where $\alpha_i = \sum_j r_{ij}$ is the order of reaction $i$. 
In the following, we focus on the analysis of non-Markovian behaviors due to extrinsic noise as reflected in scenarios $1$ and $2$.

\begin{figure}[thb]
\centering
\includegraphics[width=1\columnwidth]{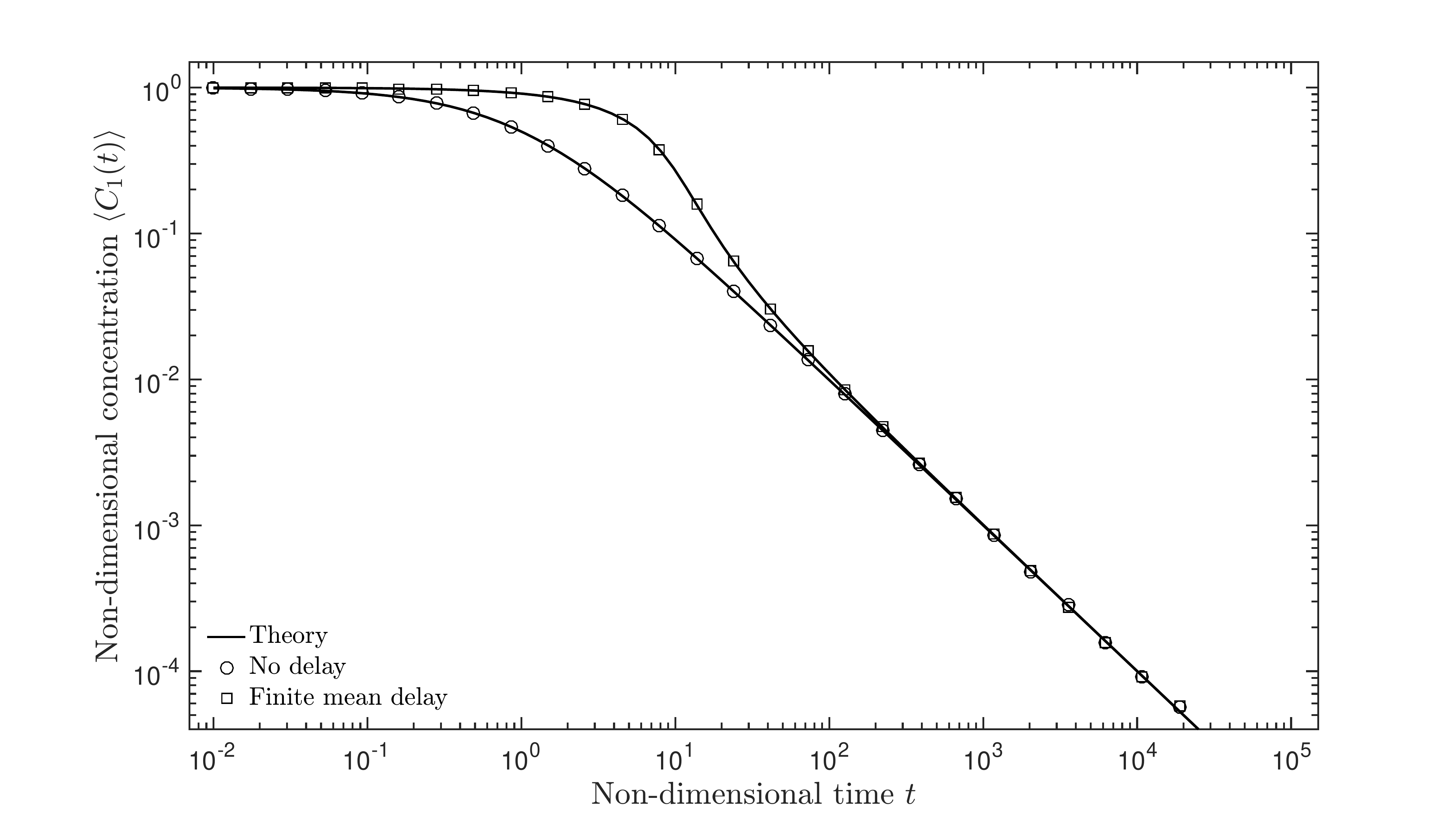}
\caption{Mean concentration for two concurrent second order annihilation reactions $\mathcal S_1+\mathcal S_2 \to \varnothing$ with exponential intrinsic waiting times, without delay and with finite-mean global delay (scenario 1). The macroscopic reaction rates are $\kappa^C_1 = 0.3$ and $\kappa^C_2 = 0.7$, and the macroscopic mean delay is $\mu^C = 10$.
Simulations (symbols) are single realizations with $n_0 = 10^6$. Time is non-dimensionalized by $t_r = 1 / [(\kappa^C_1 + \kappa^C_2) c_0]$ and concentration by $c_0$.}
\label{fi::finite_mean}
\end{figure}

\paragraph{Global delay: Scenario 1 --} We consider first a finite-mean delay with $\langle \tau^g \rangle = \mu$.  For $\lambda \ll \mu^{-1}$
we may write $\tilde\psi^g \approx 1 - \mu \lambda$. Thus, we obtain together with the exponential form of the $\phi_i^r$ given above the following approximation for the memory functions at $t \gg \mu$ (see Appendix~\ref{a::1}),
%
\begin{align}
	M_i(t;\bm{n}) &= \frac{\kappa_i h_i(\bm{n})}{1+ \mu \sum_\ell \kappa_\ell h_\ell(\bm{n})} \delta(t) \;.
\end{align}
%
The kinetic rate laws obtained from~\eqref{rate:eq} describe generalized Michaelis--Menten kinetics,
%
\begin{align}
	\partial_t \langle \bm{C} \rangle &= \sum_i \bm{s}_i \frac{ \kappa^C_i \prod_j \langle C_j \rangle^{r_{ij}}}{1+ \mu^C \sum_k \kappa^C_k \prod_\ell \langle C_\ell \rangle^{r_{k\ell}}},
\end{align}
%
where the macroscopic mean global delay $\mu^C=n_0\mu$.  Figure~\ref{fi::finite_mean} illustrates the results discussed up to here for irreversible second-order reactions $\mathcal S_1 + \mathcal S_2 \rightarrow \varnothing$ with equal initial concentrations $c_0$ for both species.

\paragraph{Global delay: Scenario 2 --} The memory functions for scenario $2$ are given by $\tilde M_i = \lambda h_i \kappa_i /[\lambda + \gamma(1 - \tilde \psi_0^g)]$. 
We consider a heavy tailed single-event delay PDF $\psi_0^g \sim t^{-1-\beta}$, such that $\tilde\psi^g_0(\lambda) \approx 1 - (\mu \lambda)^\beta$ for $\lambda \ll \mu^{-1}$. Here $\mu$ is a characteristic timescale and $0<\beta<1$. Note that such a delay is a parsimonious model for infinite-mean random variables due to the generalized central limit theorem~\cite{feller2008introduction2}. This leads to the approximate memory function
$\tilde M_i = h_i \kappa_i (t_w\lambda)^{1-\beta}$ for $t\gg t_w$ and $t \gg \mu$ (see Appendix~\ref{a::2}), where we defined the effective delay timescale 
$t_w = (\gamma\mu^\beta)^{-1/(1-\beta)}$. 
The resulting rate laws are time non-local and can be expressed in terms of fractional-in-time evolution equations,
%
\begin{equation}
\label{eq::ratelaw_frac}
	\partial_t \langle \bm{C} \rangle = \sum_i \bm{s}_i \kappa^C_i t_w^{1-\beta} \partial^{1-\beta}_t \langle \prod_j C_j^{r_{ij}} \rangle.
\end{equation}
%
Unlike for the case of finite mean delay, here $\langle \prod_j C_j^{r_{ij}}\rangle \neq  \prod_j \langle C_j\rangle^{r_{ij}}$ in the thermodynamic limit of infinite particle numbers, expressing the impact of local non-equilibrium. The ensemble average concentrations and their moments can be obtained by subordination~\cite{feller2008introduction2,benson2009simple} from the solutions of the corresponding well-mixed problem, which satisfy~\eqref{rate:eq}, see Appendix~\ref{a::2}. The behavior in single realizations of the chemical system is different from the ensemble behavior because large delay events with no change in concentration dominate. In this sense, while the intrinsic reaction conditions are the same in each realization, the global reaction behaviors are different, and thus particles in different realizations are not statistically equivalent. The system is weakly ergodicity breaking~\cite{bel2005weak,klages2008anomalous}, which is a common characteristic of anomalous transport in heterogeneous environments.

\begin{figure}[thb]
\centering
\includegraphics[width=1\columnwidth]{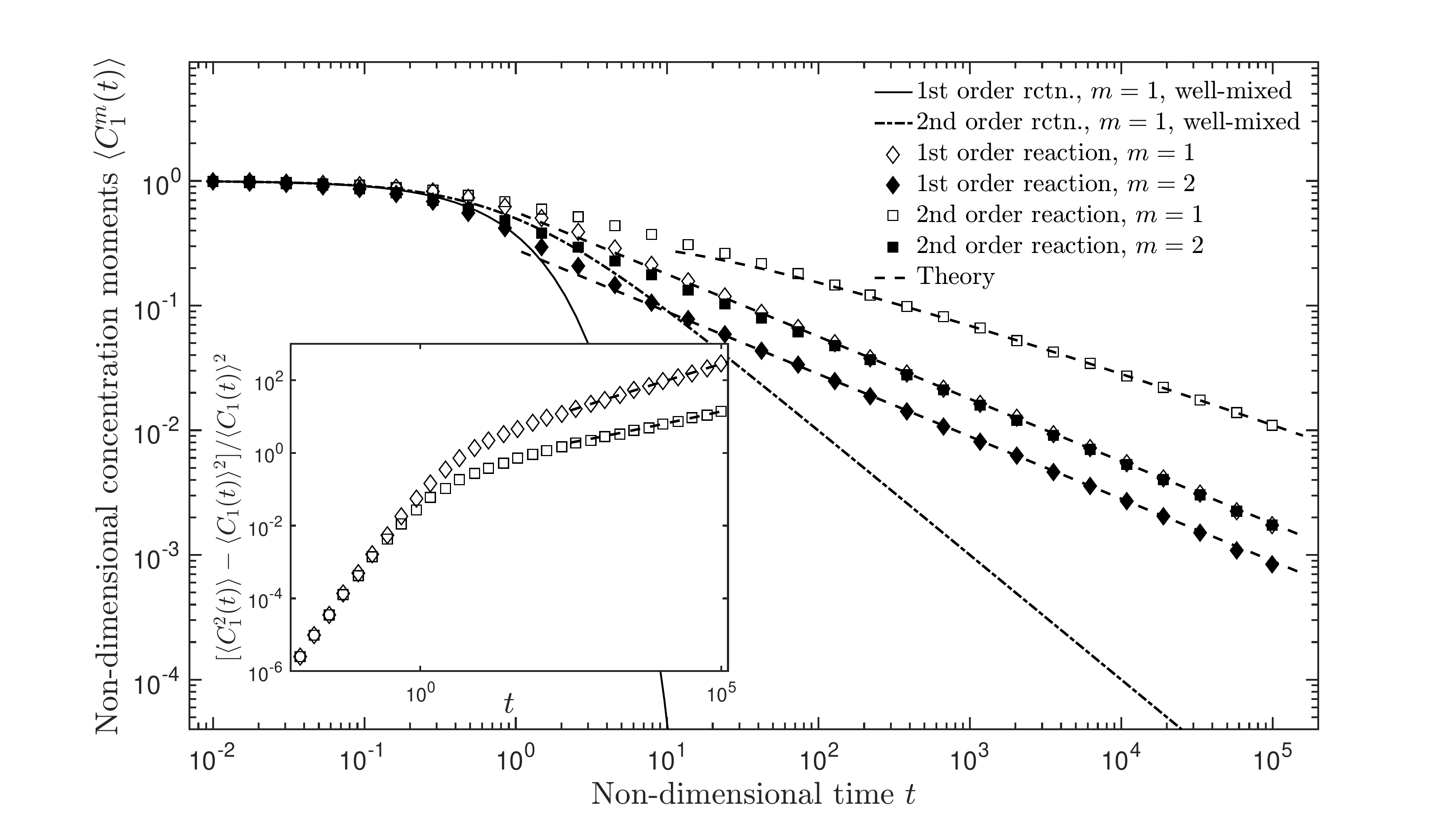}
\caption{Moments of concentration for first order $\mathcal S_1 \to \varnothing$ and second order $\mathcal S_1+\mathcal S_2 \to \varnothing$ annihilation reactions with infinite-mean delay (scenario 2). The single-event delay exponent is $\beta = 1/2$, the effective delay timescale is $t_w = 1$, and the rate of delay events is $\gamma = 10^2$. Simulations are averaged over $10^5$ realizations with $10^6$ particles. Time is non-dimensionalized by $t_r = 1 / (\kappa^C c_0^{\alpha-1})$ and concentration by $c_0$. The inset illustrates the breakdown of the $\langle C^2 \rangle = \langle C \rangle^2$ closure induced by weak ergodicity breaking.}
\label{fi::infinite_mean}
\end{figure}

To illustrate these findings, we consider annihilation reactions of order $\alpha$, $\sum_{i=1}^{\alpha} \mathcal{S}_i \to \varnothing$, with equal initial concentration $c_0$ for all species. The long time limit of Eq.~\eqref{eq::ratelaw_frac} predicts the asymptotics $\langle C_i^\alpha(t) \rangle  \propto  t^{-\beta}$. For $\alpha = 1$, all concentration moments decay algebraically. The survival probability is dominated by the distribution of reaction delays, and given by the probability that the inter-reaction time is larger than $t$. The relative concentration variance $(\langle C_i^2 \rangle - \langle C_i \rangle^2)/\langle C_i \rangle^2$ increases as $t^{\beta}$. For $\alpha = 2$, a reaction event corresponds to the annihilation of a pair, and this in turn is dictated by the delay times. This means that pair survival is governed by the delay time distribution. The mean concentration, on the other hand, behaves asymptotically as $\langle C_i \rangle \propto t^{-\beta} \ln(t)$, see Appendix~\ref{a::2}. Thus, the relative variance behaves as $t^{\beta} / \ln(t)^2$. This type of behaviors is characteristic of concentration fluctuations in random media under anomalous transport~\cite{Eisenberg:1994,CTRW2009}. Figure~\ref{fi::infinite_mean} shows the evolution of the mean and mean squared concentrations for $\alpha = 1$ and $2$.

\paragraph{Conclusions --}
We have proposed a CTRW approach for chemical reactions under non-ideal conditions which relaxes the fundamental assumptions of classical KMC methods, namely those of exponential inter-reaction times and statistical equivalence of all particles. The resulting chemical CTRW is inhomogeneous in that its evolution depends on the system state. This is a direct consequence of the dependence of the reaction waiting times on the particle numbers intrinsic to KMC.   
The global delay approach describes the impact of extrinsic noise on the reaction dynamics. It may not be applicable directly to situations in which the delay is reaction-dependent because the chemical CTRW~\eqref{eq::ctrw} implies that the delay conditions are reset after the reaction fires. The work of~\cite{brett2013stochastic} provides a framework for dealing with reaction-specific delays, although it requires \textit{ad hoc} identification of different orders of reaction firing.
In conclusion, the proposed chemical CTRW provides an approach to account for the impact of ambient fluctuations, which may open new ways of understanding and modeling reaction phenomena under non-ideal conditions. It derives generalized Michaelis--Menten kinetics as a result of finite-mean random delay, and time-nonlocal kinetic rate laws for heavy-tailed delay time distributions. 


\begin{acknowledgments}
The authors acknowledge the support of the European Research Council (ERC) through the project MHetScale (617511).
\end{acknowledgments}

%

\clearpage

\onecolumngrid

\appendix

\setcounter{secnumdepth}{2}

\section{Inter-reaction waiting times}
\label{a::waiting}

Here we provide some details on the derivation of Eqs.~(2), (7) and
(8). First, let us address the problem of obtaining the joint
density $\phi_i^r$ for reaction $i$ to occur after the reaction
waiting time $t$, where  the single-reaction waiting times are
distributed according to $\psi^r_i$. In this discussion we disregard
global delay; its inclusion is discussed in the main text. This
inter-reaction waiting time is $\tau^r(\bm n)$, given, by definition, by the minimum
waiting time to any reaction. Since the state is fixed until the next
reaction occurs, we have
%
\begin{equation}
	\tau^r(\bm{n}) = \min\{\tau^r_{(i)}(\bm{n}) \mid i = 1,\ldots,m_r\} \;,
\end{equation}
where $\tau^r_{(i)}(\bm{n})$ is the waiting time of reaction $i$. Thus, $\phi_i^r$ is given by
%
\begin{equation}
\begin{aligned}
\phi^r_i(t;\bm{n}) &= \langle \delta[t-\tau^r_{(i)}(\bm{n})] \Theta[\min(\tau_{(\ell)}^r|\ell\neq i) - t]\rangle \;,\\
&= \langle \delta[t-\tau^r_{(i)}(\bm{n})] \prod_{\ell\neq i} \Theta[\tau^r_{(\ell)}(\bm{n})-t]\rangle \;,
\end{aligned}
\end{equation}
%
where $\Theta(\cdot)$ is the Heaviside step function. The reaction waiting
times $\tau_{(i)}^r$ are mutually independent, and thus
%
\begin{equation}
\begin{aligned}
  \phi^r_i(t;\bm{n}) &= \int_0^\infty \der t' \,
\psi^r_i(t';\bm{n}) \delta(t-t') \prod_{\ell\neq i}\int_0^\infty
\der t_\ell \,\psi^r_\ell(t_\ell;\bm{n}) \Theta(t_\ell-t) \;,\\
& = \psi^r_i(t;\bm{n}) \prod_{\ell\neq i}\int_t^\infty \der t_\ell \, \psi^r_\ell(t_\ell;\bm{n}) \;.
\label{5}
\end{aligned}
\end{equation}

The expression for the PDF $\psi^r_i$ of reaction waiting times $\tau_i^r$ is obtained in the same way. The number of possible reaction events of reaction $i$ is given by $h_i(\bm n)$, which counts all possible combinations of particles and thus depends on the state $\bm n$. Each of the possible $h_i$ events is characterized by an intrinsic reaction waiting time $\theta$, which is distributed according to $p_i$. The reaction waiting time is given by
%
\begin{equation}
\tau_i^r(\bm{n}) = \min\{\theta_\ell|\ell = 1,\dots,h_i(\bm{n})\} \;.
\end{equation}
%
Its PDF is thus given by
%
\begin{equation}
\begin{aligned}
  \psi_i^r(t;\bm n) &= \langle \delta[t - \min(\theta_\ell|\ell = 1,\dots,h_i(\bm n)]  \rangle
  = \prod_{\ell = 1}^{h_i(\bm n)} \int\limits_0^\infty \der t_\ell \, p_i(t_\ell) \delta[t - \min\{\theta_\ell\}]\;,\\
                    &=                
                      \label{8}
    h_i(\bm{n}) p_i(t) \left[\int\limits_t^\infty \der t' p_i(t') \right]^{h_i-1} \;.
\end{aligned}
\end{equation}
%
The latter can also be written as
%
\begin{align}
  \label{9}
\psi_i^r(t;\bm{n}) = - \partial_t \left[\int\limits_t^\infty \der t' p_i(t') \right]^{h_i(\bm{n})} \;. 
\end{align}
%
Using~\eqref{8} and~\eqref{9} in~\eqref{5} gives
%
\begin{equation}
\begin{aligned}
  \phi_i^r(t;\bm n) &=
                      h_i(\bm n) p_i(t) \left[\int\limits_t^\infty \der t' p_i(t') \right]^{h_i(\bm n) -1} \prod_{\ell\neq i} \left[\int\limits_t^\infty \der t' p_\ell(t') \right]^{h_\ell(\bm n)} \;,\\
                    &=
                      \frac{h_i(\bm n) p_i(t)}{\int\limits_t^\infty \der t' p_i(t')} \prod_{\ell = 1}^{m_r} \left[\int\limits_t^\infty \der t' p_\ell(t') \right]^{h_\ell(\bm n)} \;.
\end{aligned}
\end{equation}
%
\section{Generalized chemical master equation}
\label{a::meq}
The recursion relations 
%
\begin{align}
\label{eq::ctrw_a}
	\bm{N}_{k+1} = \bm{N}_k + \bm{s}_{r_k} \;, \quad
	\qquad T_{k+1} = T_k + \tau_{k}
\end{align}
%
for the state vector $\bm N_k$ and time $T_k$ describe an inhomogeneous CTRW process. Thus, we now adapt the CTRW formalism in order to derive the corresponding generalized chemical master equation. We now wish to describe the evolution of particle numbers in time $t$ rather than $k$. Thus, consider the renewal process $K_t$, which describes the number of steps as a function of time. We write $\bm{N}(t)=\bm{N}_{K_t}$. The process $K_t$ is the adjoint of the process $T_k$, which describes the time elapsed after $k$ reaction steps. We have $T_{K_t}=t$, and $K_t = \sup\{k\mid T_k < t\}$. We can now write for the probability of $\bm{N}(t)=\bm{n}$
%
\begin{equation}
	P(\bm{n},t) = \langle \delta_{\bm{n},\bm{N}_{K_t}} \rangle \;,
\end{equation}
%
where the angular brackets denote the average over all realizations of the waiting time process $\tau_k$, and $\delta_{i,j}$ is the Kronecker delta. Using a partition of unity, we write
%
\begin{equation}
	P(\bm{n},t) = \sum_{k = 0}^\infty \langle \delta_{\bm{n},\bm{N}_k} \delta_{k, K_t} \rangle \;.
\end{equation}
%
The goal now is to split the average. This can be done by noticing that 
%
\begin{align}
\delta_{k, K_t} = \mathbb I(T_k \leq t < T_{k+1}) \;,
\end{align}
%
where $\mathbb I(\cdot)$ is an indicator function which is $1$ if the argument is true and $0$ otherwise. 
Thus, introducing another partition of unity, we obtain 
%
\begin{equation}
P(\bm{n},t) = \int_0^\infty \der t' \, \sum_{k = 0}^\infty \langle \delta_{\bm{n},\bm{N}_k} \delta(T_k-t') 
\mathbb I(0 \leq t - t' < \tau_{k})\rangle \;.
\end{equation}
%
Since the waiting times $\tau_k$ are independent we obtain
%
\begin{equation}
\label{eq::meq_intermediate}
	P(\bm{n},t) = \int_0^t \der t' \,  \sum_{k = 0}^\infty R_k(\bm{n},t') \sum_{i = 1}^{m_r} \int_{t-t'}^\infty \der t'' \, \phi_i(t'';\bm{n}) \;,
\end{equation}
where
%
\begin{equation}
\label{eq::ctrw_r}
	R_k(\bm{n},t) = \langle \delta_{\bm{n},\bm{N}_k} \delta(T_k-t) \rangle
\end{equation}
%
is the joint density of arriving at $\bm{n}$ at time $t$ after $k$ reaction steps. It follows that $R(\bm n,t) = \sum_{k = 0}^\infty R_k(\bm{n},t)$ is the probability per time of arriving at $\bm{n}$ at time $t$ after any number of reaction steps. Thus, Eq.~\eqref{eq::meq_intermediate} has a clear physical interpretation: The probability of finding $\bm{n}$ at time $t$ is given by the probability density of having arrived at an earlier time $t'$, and then not reacting the remaining time $t-t'$, integrated over any arrival time $t'$. Note that~\eqref{eq::ctrw_a} is a Markov process in reaction step numbers $k$ and $R_k(\bm n, t)$ is its density. Thus, $R_k(\bm n,t)$ fulfills the Chapman-Kolmogorov equation
%
\begin{align}
R_{k+1}(\bm{n},t) &= \int_0^t \der t' \, \sum_{i=1}^{m_r} R_k(\bm{n}-\bm{s}_i,t') \phi_i(t-t';\bm{n}-\bm{s}_i) \;.
\end{align}
%
Note that $R_0(\bm{n},t) = \langle \delta_{\bm{n},\bm{N}_0} \delta(T_0-t) \rangle = P(\bm{n},0) \delta(t)$, where $T_0 = 0$ was taken as zero without loss of generality. If the initial condition is deterministic, one has also $P(n,0) = \delta_{n,\bm{n}_0}$ for some given initial particle numbers $\bm{n}_0$. 

Due to the convolutions, Eqs.~\eqref{eq::meq_intermediate} and \eqref{eq::ctrw_r} form a system that is most easily solved in Laplace space, where we have
%
\begin{equation}
\begin{aligned}
	&\tilde R(\bm{n},\lambda) = P(\bm{n},0) + \sum_{i=1}^{m_r} \tilde R(\bm{n}-\bm{s}_i,\lambda) \tilde \phi_i(\lambda;\bm{n}-\bm{s}_i) \;, \\
	&\tilde P(\bm{n},\lambda) = \tilde R(\bm{n},\lambda) \frac{1 - \sum_i \tilde \phi_i(\lambda;\bm{n})}{\lambda} \;.
\end{aligned}
\end{equation}
%
It is possible to solve this system for $\tilde P$ through algebraic manipulations alone, giving
%
\begin{equation}
	\lambda \tilde P(\bm{n},\lambda) = P(\bm{n},0) + \sum_{i=1}^{m_r} \left( \prod_{j=1}^{m_s} \mathbb{E}^{-s_{ij}} - 1\right) \tilde P(\bm{n},\lambda) \tilde M_i(\lambda;\bm{n}) \;,
\end{equation}
where $\tilde M_i(\lambda;\bm{n})$ is given by Eq.~(5). Recognizing the products $\tilde P \tilde M_i$ as corresponding to convolutions in the time domain, Laplace inversion leads directly to the generalized master equation~(4).

\section{Chemical rate laws}
\label{a::rate_laws}

The generalized chemical master equation~(4) is exact, i.e., it involves no approximations given our conceptualization of the problem. However, it is sometimes convenient to describe the macroscopic behavior of a system directly. Specifically, the (ensemble) average concentration is often a quantity of interest, and may be described by a simpler, ordinary (integro-)differential equation rather than the full master equation. We thus develop here an equation for the first ensemble moment of the probability distribution of particle numbers $P$, valid at large particle numbers.

First, the step operators in Eq.~(4) may be approximated by derivatives in the following way. Using the definition of the macroscopic concentration $\bm{C}$ given in the main text, and defining also $P^C(\bm{c},t) = n_0 P(n_0 \bm{c},t)$ and $M^C_i(t;\bm{c})= M_i(t;n_0 \bm{c}) / n_0$, we can write
%
\begin{equation}
	\left(\prod_j \mathbb{E}_j^{-s_{ij}}-1\right) P(\bm{n},t') M_i(t-t';\bm{n})
		\approx - \frac{\bm{s}_{i}}{n_0} \cdot \nabla P^C(\bm{c},t') M^C_i(t-t';\bm{c}) \;.
\end{equation}
Also, for large particle numbers, we have for the sum over particle numbers (to be understood component by component) $\sum_{\bm{n}} \approx n_0 \int \der\bm{c}$. Thus, multiplying Eq.~(4) by, and summing over, $\bm{n}$, using the above approximations, and integrating over $\bm{c}$ by parts, we arrive at Eq.~(6).
\section{Generalized Gillespie algorithm}
\label{a::gillespie}
As discussed in the main text, the intrinsic reaction waiting times should be exponentially
distributed, $p_i(t) = \kappa_i \exp(-\kappa_i t)$, which recovers the
classical Gillespie algorithm in the absence of global delay. From Eq. (7) in the main text we then obtain
%
\begin{align}
\psi_i^r(t;\bm n) = h_i(\bm n) \kappa_i \exp[-\kappa_i h_i(\bm n) t] \;,
\end{align}
%
and from Eq. (8) in the main text 
%
\begin{align}
\label{s20}
\phi_i^r(t;\bm n) = h_i(\bm n) \kappa_i \exp\left[-\sum_\ell
  \kappa_\ell h_\ell(\bm n) t \right] \;.
\end{align}
%
The probability $\rho_i$ that reaction $i$ occurs is given by marginalization
of the latter over $t$ as 
%
\begin{align}
\label{rhoi}
\rho_i(\bm n) = \frac{h_i(\bm n) \kappa_i}{\sum_\ell
  \kappa_\ell h_\ell(\bm n)} \;.
\end{align}
%
The PDF of reaction waiting times $\phi^r_{|i}$, given that reaction $i$ occurs, is accordingly given
by
%
\begin{align}
\label{s22}
\phi^r_{|i}(t;\bm n) = \sum_\ell
  \kappa_\ell h_\ell(\bm n) \exp\left[-\sum_k
  \kappa_k h_k(\bm n) t \right] \;,
\end{align}
%
which is independent of $i$. 
Based on these two distributions, we can now describe the algorithm,
one step of which may be summarized as follows: 
\begin{enumerate}

\item Generate a random integer $i$ according to~\eqref{rhoi}. 

\item Generate the reaction waiting time $\tau^r$ according
  to~\eqref{s22}. 

\item Generate the global delay time $\tau^{g}$:

\begin{enumerate}

\item scenario 1: according to $\psi^g$. 

\item scenario 2: generate a random variable $\eta$ according to a
  Possion distribution with mean $\gamma \tau^r$, where $\gamma$ is
  the (reaction-independent) rate at which delay events occur; generate a
  series of $\eta$ random variables $\{\vartheta^g_k\}_{k=1}^\eta$ according to $\psi_0^g$; determine
  the global delay as $\tau^g = \sum_{k=1}^\eta \vartheta_k^g$. 
\end{enumerate}

\item Increment time by $\tau^r + \tau^g$.

\item  Change the system state according to reaction $i$. 

\end{enumerate}
These procedures are to be repeated until a certain condition is met,
such as a certain maximum time being exceeded. 

Note that this algorithm may serve two related but slightly different purposes: (i) Directly simulate the dynamics of a reactive system for which the reaction waiting times are known in the context of a CTRW; and (ii) Numerically integrate a known generalized master equation of the form~(4) by constructing appropriate reactions.

\section{Scenario 1: Independent global delay}
\label{a::1}

First, let us consider fully independent global delay, distributed according to $\psi^g$. From Eq.~(5), and using Eq.~\eqref{s20},
%
\begin{equation}
	\tilde M_i(\lambda;\bm{n}) = \frac{\lambda \kappa_i h_i(\bm n)
        \tilde \psi^g(\lambda)}{\lambda + [1 - \tilde \psi^g(\lambda)]
        \sum_\ell \kappa_\ell h_\ell(\bm n)} \;.
\end{equation}
For a finite mean global delay such that  $\langle \tau^g \rangle =
\mu < \infty$,  we approximate $\tilde \psi^g(\lambda) \approx 1 - \mu
\lambda$ for $\lambda \ll \mu^{-1}$, which gives 
%
\begin{equation}
	\tilde M_i(\lambda;\bm{n}) \approx \frac{\kappa_i h_i(\bm{n})}{1+\mu\sum_\ell \kappa_\ell h_\ell(\bm{n})} \;.
\end{equation}
%
\section{Scenario 2: Global delay as a compound Poisson process}
\label{a::2}
\subsection{Memory function}
For exponentially distributed intrinsic waiting times we obtain the exact result
%
\begin{equation}
\label{eq::memory_compound_exp}
	\tilde M_i(\lambda;\bm{n}) = \frac{h_i(\bm{n}) \kappa_i  \lambda}{\lambda+\gamma[1-\tilde\psi^g_0(\lambda)]} \;.
\end{equation}
%
Now assume that the trapping times have infinite mean and some
characteristic time scale $\mu$, such that
$\tilde\psi^g_0(\lambda)\approx1-(\mu\lambda)^\beta$ for
$\lambda\ll\mu^{-1}$, with $0<\beta<1$.  For $\lambda\ll t_w^{-1} = (\gamma\mu^\beta)^{1/(1-\beta)}$ we find
%
\begin{equation}
	\tilde M_i(\lambda;\bm{n}) \approx h_i(\bm{n}) \kappa_i (t_w \lambda)^{1-\beta} \;.
\end{equation}

%
\subsection{Moment asymptotics for infinite-mean delay}

In order to obtain the asymptotic behavior for concentration moments arising from Eq.~(12), we first consider its Laplace transform,
%
\begin{equation}
	\lambda \langle \tilde{\bm{C}} \rangle = \bm{C}_0 + \sum_i
        \bm{s}_i \kappa^C_i (t_w \lambda)^{1-\beta} \mathcal L\{ \langle \prod_j C_j^{r_{ij}} \rangle \} \;,
\end{equation}
where $\mathcal L\{\cdot\}$ denotes the Laplace transform. 
Now, for an order $\alpha$ annihilation reaction $\sum_{i=1}^{\alpha} \mathcal{S}_i \to \varnothing$ with equal initial concentrations $c_0$, we have $C_i(t) = C(t)$ for all species $i$ and all times $t$ due to the initial condition and the reaction stoichiometry, and we obtain
%
\begin{equation}
	\lambda \langle \tilde{C} \rangle = c_0 - \kappa^C (t_w
        \lambda)^{1-\beta} \mathcal L\{ \langle C^\alpha \rangle \} \;.
\end{equation}
For small $\lambda$ (late times), the dominant terms in this equation
are the two on the right hand side. Solving for $\mathcal L\{ \langle C^\alpha \rangle \}$ and inverting the Laplace transform yields directly
%
\begin{equation}
\label{eq::moments_delay}
	\frac{\langle C^\alpha(t) \rangle}{c_0^\alpha} \approx \frac{1}{\Gamma(1-\beta)} \left(\frac{t_r}{t_w}\right)^{1-\beta}\left(\frac{t}{t_r}\right)^{-\beta} \;,
\end{equation}
where $t_r = 1/(\kappa^C c_0^{\alpha-1})$, $t_w = (\gamma\mu^\beta)^{-1/(1-\beta)}$, and $\Gamma$ is the Gamma function, for $t \gg \mu$, $t_w$, $t_r (t_r/t_w)^{(1-\beta)/\beta}$ \;.

\subsection{Subordination approach}
\label{a::subordination}
The generalized chemical master equation for exponential intrinsic
reaction times reads in Laplace space as
%
\begin{align}
\label{gmeso}
\lambda \tilde P(\bm{n},\lambda) = \tilde P(\bm n,0) + \sum_i \left(\prod_j \mathbb{E}_j^{-s_{ij}}-1\right) 
\tilde P(\bm{n},\lambda) \frac{h_i(\bm{n}) \kappa_i  \lambda}{\lambda+\gamma[1-\tilde\psi^g_0(\lambda)]} \;.
\end{align}
%
Specifically, for
$\tilde \psi_0^g \approx 1 - (\mu \lambda)^\beta$, we obtain 
%
\begin{align}
\label{gmesolevy}
\lambda \tilde P(\bm{n},\lambda) \approx \tilde P(\bm n,0) + \sum_i \left(\prod_j \mathbb{E}_j^{-s_{ij}}-1\right) 
\tilde P(\bm{n},\lambda) \frac{h_i(\bm{n}) \kappa_i}{1+ (t_w \lambda)^{\beta - 1}} \;,
\end{align}
%
where $t_w = (\gamma \mu^\beta)^{1/(\beta-1)}$. Note that for a fixed finite $t_w$, this equation is exact for all $\lambda$ in the scaling limit $\mu \to 0$, $\gamma \to \infty$.
We can write~\eqref{gmesolevy} as 
%
\begin{equation}
\begin{aligned}
\label{gmesolevy2}
[\lambda + (t_w \lambda)^{\beta}/t_w] \tilde P(\bm{n},\lambda) &=
[1+ (t_w \lambda)^{\beta - 1}]\tilde P(\bm n,0)\\
&\quad+\sum_{i=1}^{m_r} \left(\prod_j \mathbb{E}_j^{-s_{ij}}-1\right) 
\tilde P(\bm{n},\lambda) h_i(\bm{n}) \kappa_i \;.
\end{aligned}
\end{equation}
%
It has the same form as the Laplace transform of the chemical master for $P_{wm}$ in the well- mixed scenario,
%
\begin{equation}
\begin{aligned}
\label{mep}
\lambda \tilde P_{wm}(\bm{n},t) = P_{wm}(\bm n,0) +\left(\prod_j \mathbb{E}_j^{-s_{ij}}-1\right) 
\tilde P_{wm}(\bm{n},\lambda) h_i(\bm n) \kappa_i \;,
\end{aligned}
\end{equation}
%
where the subscript $wm$ denotes \textit{well mixed}. Thus, $\tilde P$ can be expressed in terms of $\tilde P_{wm}$ as 
%
\begin{align}
\label{ppwm}
\tilde P(\bm n,\lambda) =  [1 + (t_w\lambda)^{\beta-1}] 
\tilde P_{wm}[\bm n,\lambda + (t_w\lambda)^\beta / t_w] \;.
\end{align}
%

The latter can also be obtained by subordination with the time process 
%
\begin{align}
\der T(u) = \der u + t_w(\der u / t_w)^{1/\beta} \xi(u) \;,
\end{align}
%
where the $\xi(u)$ are independent unit (i.e., with unit characteristic time) L\'evy $\beta$-stable random variables. We then obtain for $P$ 
%
\begin{align}
\label{eqso}
P(\bm n,t) = \int\limits_0^\infty \der u \, P_{wm}(\bm n,u) h(u,t) \;,
\end{align}
%
where $h(u,t) = \delta[u - U(t)]$, with $U(t) = \max[u|T(u) \leq t]$.  
We note now that 
%
\begin{equation}
\int\limits_u^\infty \der u' \, h(u',t) = \int\limits_0^t dt' \, l(t',u) \;,
\end{equation}
%
where $l(t,u) = \langle \delta(t - T(u)) \rangle$. We obtain for their Laplace transforms 
%
\begin{align}
\tilde h(u,\lambda) = - \lambda^{-1} \partial_u \tilde l(\lambda,u) \;. 
\end{align}
%
The Laplace transform of $l(t,u)$ is 
%
\begin{equation}
	\tilde l(\lambda,u) = e^{-[\lambda + (t_w \lambda)^\beta/t_w]u} \;,
\end{equation}
%
and therefore
%
\begin{equation}
	\tilde h(u,\lambda) = [1 +  (t_w\lambda)^{\beta-1}] e^{-[\lambda + (t_w\lambda)^{\beta}/t_w]u} \;.
\end{equation}
%
Using the latter in the Laplace transform of~\eqref{eqso} gives for $\tilde P$
%
\begin{align}
\tilde P(\bm n,\lambda) = [1 +  (t_w\lambda)^{\beta-1}]
  \int\limits_0^\infty \der u P_{wm}(\bm n,u) e^{-[\lambda +
  (t_w\lambda)^{\beta}/t_w]u} \;,
\end{align}
%
which is equivalent to~\eqref{ppwm}. 

Thus, we may calculate the ensemble average of $\tilde{\bm{C}}$ as
%
\begin{equation}
\label{eq::average_subordination}
	\langle \tilde{\bm{C}}(\lambda) \rangle = [1 +  (t_w\lambda)^{\beta-1}] 
\langle \tilde{\bm{C}}_{wm}[\lambda + (t_w\lambda)^{\beta}/t_w]\rangle \;.
\end{equation}
%

\subsubsection{Annihilation Reaction $\alpha = 2$}

We now compute $\tilde{\bm{C}}$ for the reaction $\mathcal S_1 + \mathcal S_2 \to \varnothing$ with equal initial conditions $c_0$ for both species. Due to the stoichiometry and the initial condition, we have $C_1(t) = C_2(t) = C(t)$ for all times $t\geqslant0$. The well-mixed solution is also the same for each species, $\langle C_{wm}(t) \rangle = c_0 / ( 1 + t / t_r )$, with $t_r = 1 / (\kappa^C c_0)$. Taking the Laplace transform of the well-mixed solution and using Eq.~\eqref{eq::average_subordination} leads to
%
\begin{equation}
	\frac{\langle \tilde{C}(\lambda) \rangle}{c_0} = -[1 + (t_w \lambda)^{\beta-1}] t_re^{t_r \lambda}\mathrm{Ei}[t_r(\lambda + (t_w \lambda)^\beta/t_w)]  \;,
\end{equation}
where $\mathrm{Ei}(x) = \int_{-x}^{\infty} \der y \, e^{-y} / y$ is an exponential integral.

To determine the late-time behavior, we consider the following expansion around $x=0$,
%
\begin{equation}
	\mathrm{Ei}(x) = \gamma_E + \ln(|x|) + \mathcal{O}(x) \;,
\end{equation}
where $\gamma_E = -\digamma(1)$ is the Euler--Mascheroni constant, with $\digamma$ the digamma function. This leads to
%
\begin{equation}
	\frac{\langle \tilde{C}(\lambda) \rangle}{c_0} \approx
		-\frac{t_r \beta}{(t_w \lambda)^{1-\beta}} \ln\left[ \left(\frac{e^{\gamma_E} t_r}{t_w}\right)^{1/\beta} t_w \lambda \right] \;,
\end{equation}
for $\lambda \ll t_r^{-1}, t_w^{-1}, \mu^{-1}, t_r^{-1}
(t_r/t_w)^{-(1-\beta)/\beta}$. Using the inverse Laplace
transform $\mathcal{L}^{-1}\{{\ln \lambda}/{\lambda}\} = -(\gamma_E + \ln t)$,
we obtain
%
\begin{equation}
	\frac{\langle C(t) \rangle}{c_0} \approx
		\frac{t_r}{t_w^{1-\beta}}\beta\partial_t^\beta \ln\left[ \left(\frac{t_w}{e^{\gamma_E(1-\beta)}t_r}\right)^{1/\beta} \frac{t}{t_w} \right] \;.
\end{equation}
The fractional derivative of the logarithm can be computed explicitly~\cite{tu1995certain}. For $0<\beta<1$,
%
\begin{equation}
	\partial_x^\beta \ln x =
		\frac{x^{-\beta}}{\Gamma(1-\beta)}[\ln x - \gamma_E - \digamma(1-\beta)] \;,
\end{equation}
and we obtain, in terms of $t/t_r$,
%
\begin{equation}
	\frac{\langle C(t) \rangle}{c_0} \approx
		\left(\frac{t_r}{t_w}\right)^{1-\beta}\left(\frac{t}{t_r}\right)^{-\beta}
		\frac{\beta}{\Gamma(1-\beta)}
		\left(
			\ln\left[\left(\frac{t_w}{t_r}\right)^{\frac{1-\beta}{\beta}}\frac{t}{t_r}\right]
			- \frac{\gamma_E}{\beta} - \digamma(1-\beta)
		\right) \;.
\end{equation}
\subsubsection{Annihilation Reaction $\alpha = 1$}

Higher order moments of concentration may also be obtained using the subordination approach for this and other reaction setups. For example, we may easily obtain all integer-order moments for a first order annihilation reaction $\mathcal S_1 \to \varnothing$. First, we consider the appropriate well-mixed solution, which in this case is given by $\langle C_{wm}(u) \rangle = c_0 e^{-\kappa^C u} = c_0 e^{-u / t_r}$. The $m$th order moment is then given by the formula
%
\begin{equation}
	\langle C_1^m(t) \rangle = \int_0^t\der u \, \langle C_{wm}(u) \rangle^m h(u,t) \;,
\end{equation}
where we have used the closure $\langle C_{wm}^m(u) \rangle = \langle C_{wm}(u) \rangle^m$, which holds for the well mixed process. Using the same approach as above, we can easily solve this integral by considering the Laplace transform, and we obtain
%
\begin{equation}
	\frac{\mathcal L\{\langle C_1^m \rangle\}}{c_0^m} = \frac{t_r + t_r(t_w\lambda)^{\beta-1}}{m + (t_r/t_w)(t_w \lambda)^{\beta} + t_r\lambda} \;,
\end{equation}
which yields, for late times $t \gg t_r, t_w, t_r (t_r/t_w)^{(1-\beta)/\beta}/m$,
%
\begin{equation}
	\frac{\langle C_1^m(t) \rangle}{c_0^m} = \frac{1}{m \Gamma(1-\beta)}\left(\frac{t_r}{t_w}\right)^{1-\beta}\left(\frac{t}{t_r}\right)^{-\beta} \;.
\end{equation}
Notice how the first order moment agrees with that obtained from Eq.~(12), see Eq.~\eqref{eq::moments_delay}.

\end{document}